\documentclass[graybox]{svmult}
\usepackage{mathptmx}       
\usepackage{helvet}         
\usepackage{courier}        
\usepackage{type1cm}        
\usepackage{makeidx}         
\usepackage{graphicx}        
\usepackage{multicol}        
\usepackage[bottom]{footmisc}
\makeindex             

\begin{document}

\title*{Interacting Supernovae: Types IIn and Ibn}

\titlerunning{Interacting Supernovae} 
\author{Nathan Smith}
\authorrunning{N.\ Smith} 
\institute{Nathan Smith \at Steward Obervatory, University of Arizona,
  933 N. Cherry Ave., Tucson, AZ 85721, USA,
  \email{nathans@as.arizona.edu}}
\maketitle


\abstract{Supernovae that show evidence of strong shock interaction
  between their ejecta and pre-existing, slower circumstellar material
  (CSM) constitute an interesting, diverse, and still poorly
  understood category of explosive transients.  The chief reason that
  they are extremely interesting is because they tell us that in a
  subset of stellar deaths, the progenitor star may become wildly
  unstable in the years, decades, or centuries before explosion. This
  is something that has not been included in standard stellar
  evolution models, but may significantly change the end product and
  yield of that evolution and complicates our attempts to map SNe to
  their progenitors.  Another reason they are interesting is because
  CSM interaction is a very efficient engine for making bright
  transients, allowing super-luminous transients to arise from normal
  SN explosion energies, and allowing transients of normal supernova
  luminosities to arise from sub-energetic explosions or low
  radioactivity yield.  CSM interaction shrouds the fast ejecta in
  bright shock emission, obscuring our normal view of the underlying
  explosion, and the radiation hydrodynamics of the interaction is
  challenging to model. The CSM interaction may also be highly
  non-spherical, perhaps linked to binary interaction in the
  progenitor system.  In some cases, these complications make it
  difficult to definitively tell the difference between a core
  collapse or thermonuclear explosion, or to discern between a
  non-terminal eruption, failed supernova, or weak supernova.  Efforts
  to uncover the physical parameters of individual events and
  connections to possible progenitor stars make this a rapidly
  evolving topic that continues to challenge paradigms of stellar
  evolution.}

\section{Introduction}
\label{sec:1}
Type~IIn superovae (SNe IIn hereafter), defined broadly as SNe that
exhibit bright and narrow Balmer lines of H in their spectra, were
recognized as a distinct class of objects relatively late compared to
other SN subtypes (see Schlegel 1990 and Filippenko 1997 for early
background).  This was spurned largely by observations of the classic
SN~IIn 1988Z and connecting it to unusual properties in some other
SNe.  Our understanding of this class and various subclasses has been
evolving continuously since then as more SNe are discovered and the
diversity of SNe~IIn grows.  The related class of SNe~Ibn, showing
narrow lines of He in the optical spectrum (and relatively weak H
lines), was clearly identified only 10 years ago following studies of
the prototype SN~2006jc (Pastorello et al.\ 20017; Foley et al.\
2007).

Interpreting observations of interacting SNe is fundamentally
different from studying ``normal'' SNe. For our purposes here,
``normal'' SNe are those cases where the observed emission comes from
a photosphere receding (in mass coordinates) back through freely
expanding SN ejecta, or radiation from the optically thin SN ejecta at
late times that are heated internally by radioactivity.  These are the
main subtypes of II-P, II-L, IIb, Ib, Ic, Ic-BL discussed elsewhere in
this volume.  SNe IIn and Ibn represent a physically very different
scenario, where the usual diagnostics and physical pictures cannot be
applied.  As a means of introduction to the topic, we begin with two
warnings that should be heeded when thinking of SNe IIn and related
subtypes of interacting SNe:

\begin{svgraybox}
  Warning 1.  The main engine is conversion of kinetic energy to light
  in a shock at the outside boundary of the SN. It is not
  shock-deposited energy leaking out of a homologously expanding
  envelope (as in early phases of a SN~II-P), nor is it internal
  heating from radioactive decay (most SNe at peak and at late times).
  Instead, the observed radiation comes primarily from a shock
  crashing into dense CSM which - unfortunately from the point of view
  of predictive models - makes the situation complicated and
  malleable.  ``Malleable'' here means that it is hard to calculate
  predictive models because the shock luminosity can rise and fall
  somewhat arbitrarily depending on the radial density structure of
  the CSM.  One should not take the absence of normal signatures
  (broad lines in the SN ejecta photosphere or nebular emission lines)
  as an indication that these are not true SNe, because CSM
  interaction can mask these normal signatures.  Moreover, radiation
  from the shock can propagate back into the SN ejecta, changing their
  appearance.
\end{svgraybox}

\begin{svgraybox}
  Warning 2.  SNe~IIn and related categories are not really a {\it
    supernova type} (or more accurately, not an intrinsic explosion
  type), but an {\it external phenomenon} associated with CSM
  interaction.  Any type of core collapse or thermonuclear SN (or for
  that matter, any non-SN explosive outflow) can appear as a SN~IIn or
  SN~Ibn.  All that is required is fast ejecta with sufficient energy
  crashing into slower ejecta with sufficient density.  This is a
  cause of much confusion and uncertainty.
\end{svgraybox}

As a consequence of these differences compared to other SNe, we find a
very diverse range of properties in interacting SNe.  There is not a
singular evolutionary path that leads to an interacting SN, nor is
there a singular mapping of SNe IIn to a specific progenitor type.
For example, we usually associate SNe~II-P as coming from moderately
massive red supergiants of 8-20 M$_{\odot}$ (Smartt 2009), but such
assignments are not so straightforward for SNe~IIn.  It can be
difficult to tell the difference between a SN IIn that arises from the
core collapse of a very massive luminous blue variable (LBV) or
LBV-like star, and a thermonuclear explosion of a white dwarf with
dense CSM.  SNe~IIn can arise any time there is dense H-rich material
sitting around a star that explodes, which might happen in a number of
different ways.  Exploring this diversity in progenitors as well as in
CSM and explosion properties is a main emphasis of current work
because it has important implications for mass loss and instability in
the late phases of stellar evolution.

The detailed physical picture and time evolution of interacting SNe
may also be very different from normal SNe.  Instead of a more-or-less
spherical photosphere receding through outflowing ejecta, we have
radiation from a thin shell, which may start out as an optically thick
sphere, but may transition to a limb-brightened shell that is
transparent in the middle and optically thick at the edges.  We may
have underlying explosions that would otherwise be too faint to be
observed (for example, explosions producing very low radioactive
yield), and that don't contribute to the traditional populations of
extragalactic SNe, but nevertheless yield bright transients through
their CSM interaction.  Thus, we should be mindful that some SNe~IIn
may be powered by entirely new (or so far observationally
unrecognized) classes of explosions other than SNe II-P, II-L, IIb,
Ib, and Ic.  Much of this is still little explored or uncharted
territory.

For this chapter, it is also worth making a distinction of scope.
Actually, all SNe are interacting with surrounding material at some
level, because space is not a true vacuum and an interaction arises
naturally even when fast SN ejecta expand into low density ISM or
their progenitor's normal wind (this eventually forms a standard SN
remnant).  Whether this can be detected in extragalactic SNe shortly
after explosion --- and in what waveband it is detected --- depends to
a large extent on the CSM density.  For the purposes of discussion
here, we take ``Interacting SNe'' to mean those with extraordinarily
strong CSM interaction that leads to a radiative shock at early times
after explosion, producing narrow lines in the visible-wavelength
spectrum and enhanced optical continuum luminosity at early times.  We
exclude otherwise normal SNe that show signs of interaction based on
radio or X-ray emission, although it should be kept in mind that these
represent the tail of a continuum of CSM density stretching from
superluminous SNe, through more traditional SNe~IIn, and on to lower
CSM density.

\begin{figure}[b]
\includegraphics[scale=.55]{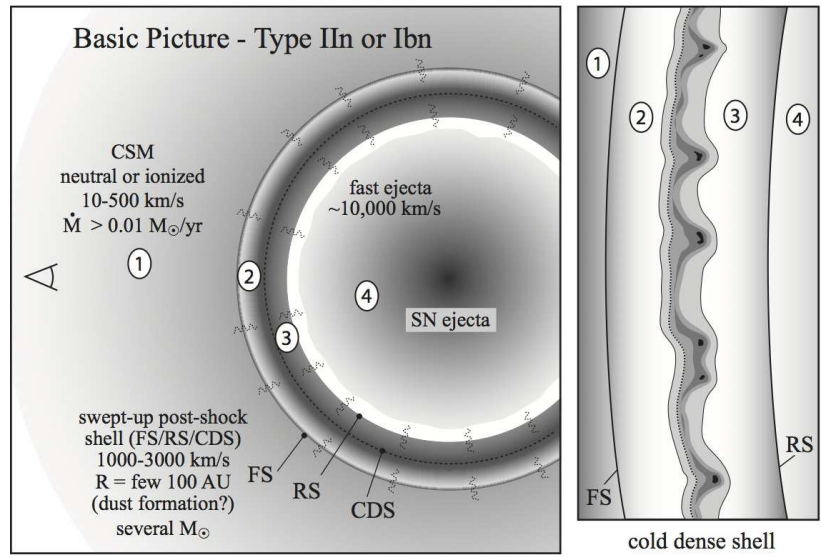}
\caption{Sketch of the basic picture in CSM interacting SNe.  Four
  different zones are noted with numbers: (1) the pre-shock CSM, (2)
  the shocked CSM, (3) the shocked SN ejecta, and (4) the freely
  expanding SN ejecta.  These zones are divided by boundaries
  corresponding to the {\color{green} forward shock}, the
  {\color{green} reverse shock}, and the {\color{green} contact
    discontinuity} between the shocked CSM and shocked ejecta where
  material cools, mixes via Rayleigh-Taylor instabilities, and piles
  up.  This is often called the {\color{green} cold dense shell (CDS)}
  in a SN IIn or Ibn.  The squiggly radial lines are meant as a
  reminder that X-rays and UV radiation generated in the shock can
  propagate out to the CSM or inward to the unshocked ejecta, changing
  the physical state of the gas there.  A zoom-in of zones 2 and 3 is
  shown at the right.  In practice, efficient radiative cooling can
  cause these two zones to collapse to very thin layers, and mixing
  can make them merge into one thin clumpy shell. This figure is
  adapted from Smith et al. (2008).
}
\label{fig:06tf}     
\end{figure}

\section{Basic Physical Picture}
\label{sec:2}

When a SN explodes inside a dense cocoon of CSM, a strong shock is
driven into the CSM, creating a basic structure as depicted in
Figure~\ref{fig:06tf}.  A forward shock (FS) is driven outward into
the CSM, and a corresponding reverse shock (RS) is driven back into
the SN ejecta.  In the simplest picture, four zones are delineated
where heated material can radiate and contribute to the observed
spectrum:

\begin{description}[Zones]
\item[Zone 1.]{The unshocked CSM outside the forward shock (photoionized).}
\item[Zone 2.]{The swept-up CSM that has been hit by the forward shock.}
\item[Zone 3.]{The decelerated SN ejecta that have encountered the reverse shock.}
\item[Zone 4.]{The freely expanding SN ejecta.}
\end{description}

In normal SNe, we only see emission from Zone 4 at visible wavelengths
as the photosphere recedes back through the freely expanding ejecta,
creating broad emission and absorption lines whose velocities decrease
smoothly with time as the emitting layer recedes through the monotonic
outflow. Zones 2 and 3 are sometimes detected in X-rays or radio
emission if the CSM is dense enough.  Zone 1 might be seen in
absorption lines, although this is rare.

In an interacting SN, any of these 4 zones can contribute strongly to
the emitted visible-wavelength spectrum.  In fact, the continuum
photosphere moves through all 4 zones as time proceeds, and their
relative contributions to line emission change with time.  The basic
picture of this interaction, which can differ widely in detail from
case to case, has been described many times in the literature (Chugai
1997, 2001; Chugai \& Danziger 1994; Chugai et al.\ 2004; Smith et
al.\ 2008, 2010a, 2015; Dessart et al.\ 2015).

At early times, radiation from the shock propagates upstream and
causes an ionized precursor.  At this stage (sometimes only lasting a
day or two, sometimes several weeks or months), the electron
scattering continuum photosphere is actually in the unshocked CSM,
obstructing our view of the shock.  The spectrum usually appears as a
smooth blue continuum with narrow emission lines that have broad
Lorentzian wings.  These wings are not caused by Doppler motion of
shocked gas, but by electron scattering as narrow line photons work
their way out of the optically thick CSM.

As the pre-shock CSM density drops at larger radii, eventually the
photosphere recedes into zones 2 and 3, and we begin to see a
transformation in the spectrum.  This usually occurs around the time
of peak luminosity.  Strong intermediate-width (few 10$^3$ km
s$^{-1}$) components of permitted lines appear, sometimes accompanied
by P Cygni absorption.  Here we are seeing radiation from the
post-shock gas.  Because of strong radiative cooling (which occurs by
definition in an interacting SN), Zones 2 and 3 can collapse into a
geometrically very thin layer, and mixing at the contact discontinuity
may cause these two zones to merge.  The standard picture is that this
cooled gas is very dense and piles up in a ``cold dense shell''
(Chugai et al.\ 2004).  This gas is continually reheated by the shock,
and emits strongly in lines like H$\alpha$ or He~{\sc i} that appear
as the strongest intermediate-width lines in the spectra of SNe IIn
and Ibn, respectively.  This line emission continues to be strong even
after the continuum photosphere recedes further, because the CDS is
reheated by X-ray and UV radiation from the FS and RS.

After peak, the continuum photosphere may recede back into Zone 4 in
the freely expanding SN ejecta.  If the CSM interaction is relatively
weak or shortlived, an observer might then see a fairly normal SN
spectrum with broad P Cygni lines at the end of the photospheric
phase, or typical signatures of SN ejecta heated by radioactive decay
in the nebular phase.  However, if CSM interaction is strong and
sustained, an observer might never actually see a normal SN ejecta
photosphere or nebular phase, and the P Cygni absorption may be filled
in.  The underlying SN (heated by the original shock deposition or
radioactive decay) may fade in a few months, whereas CSM interaction
might remain very bright for many months or years.  How long this CSM
interaction stays bright is determined mainly by the duration and
speed of the pre-SN mass loss, not by the explosion.  By the time the
CSM fades enough to become transparent, the inner SN ejecta may be too
faint to detect on their own, and the ongoing CSM interaction may
continue to dominate the spectrum at late times.  In fact, the fast SN
ejecta may be heated predominantly by radiation from the CSM
interaction shock that propagates backward into the outermost ejecta,
creating a very different time dependent spectrum than is seen in a
normal SN heated from the inside-out by radioactive decay.  One must
bear these differences in mind when interpreting SNe~IIn spectra,
especially at late phases.

Of course, if the geometry is asymmetric, any of these zones can be
seen simultaneously and at different characteristic velocities,
potentially making the interpretation quite complicated. Absorption
features may or may not be seen, depending on viewing angle. Even if
the interaction is spherically symmetric, we might see multiple zones
simultaneously because long path lengths through the limb-brightened
edges of the CDS may remain optically thick while material along our
line of sight in the CDS has already become transparent.  Very strong
and non-monotonic changes in density and velocity make this a
complicated situation that presents a significant challenge for
radiation transfer codes.  For the time being, unraveling the various
clues in the spectral evolution is something of an art form.

The total luminosity generated by CSM interaction in this picture can
be high because a radiative shock is a very efficient engine to
convert kinetic energy into visible-wavelength light.  Whereas a
normal SN II might have a total radiated energy that is 1\% of its
kinetic energy, a Type II SN can have a total radiated energy closer
to 50\% or more of the total explosion kinetic energy. The luminosity
of CSM interaction $L$ depends essentially on the rate at which CSM
mass is entering the forward shock, which of course depends mostly on
the progenitor's mass loss rate, and is typically expressed as

\begin{equation}
L = \frac{1}{2} w V_{\rm CDS}^3,
\end{equation}

\noindent where $w$ is the wind density parameter $w = 4 \pi R^2
\rho$, or $w = \dot{M} / V_{\rm CSM}$, and $V_{\rm CDS}$ is the value
for the evolving speed of the CDS.  From the intermediate-width lines
in optical spectra we can measure the value of $V_{\rm CDS}$ (be
careful not to do this at early times when electron scattering in the
CSM is determining the line width; at early times the true value of
$V_{\rm CDS}$ is hidden from the observer, and it may be much higher
than when it is seen at later times).  We can also measure $V_{\rm
  CSM}$ from the narrow component due to pre-shock gas, especially if
a narrow P Cygni profile is observed (note here that higher resolution
spectra with $R \simeq 4000$ are especially useful).  Using the
luminosity ($L$) derived from photometry (note that for $V$ and $R$
bands, the bolometric correction may be quite small if the apparent
temperature is around 6000-7000 K), we can then turn this equation
around to estimate the progenitor star's mass-loss rate

\begin{equation}
\dot{M}_{\rm CSM} = 2 \ L \ \frac{V_{\rm CSM}}{(V_{\rm CDS})^3},
\end{equation}

\noindent that occurred at some time $t$=$t_{\rm exp} (V_{\rm
  CDS}/V_{\rm CSM})$ in the past, where $t_{\rm exp}$ is the time
after explosion to which these measurements correspond (careful again,
as $t_{\rm exp}$ has its own potentially large uncertainties).

\newpage

\begin{figure}[h]
  \sidecaption
\includegraphics[scale=.6]{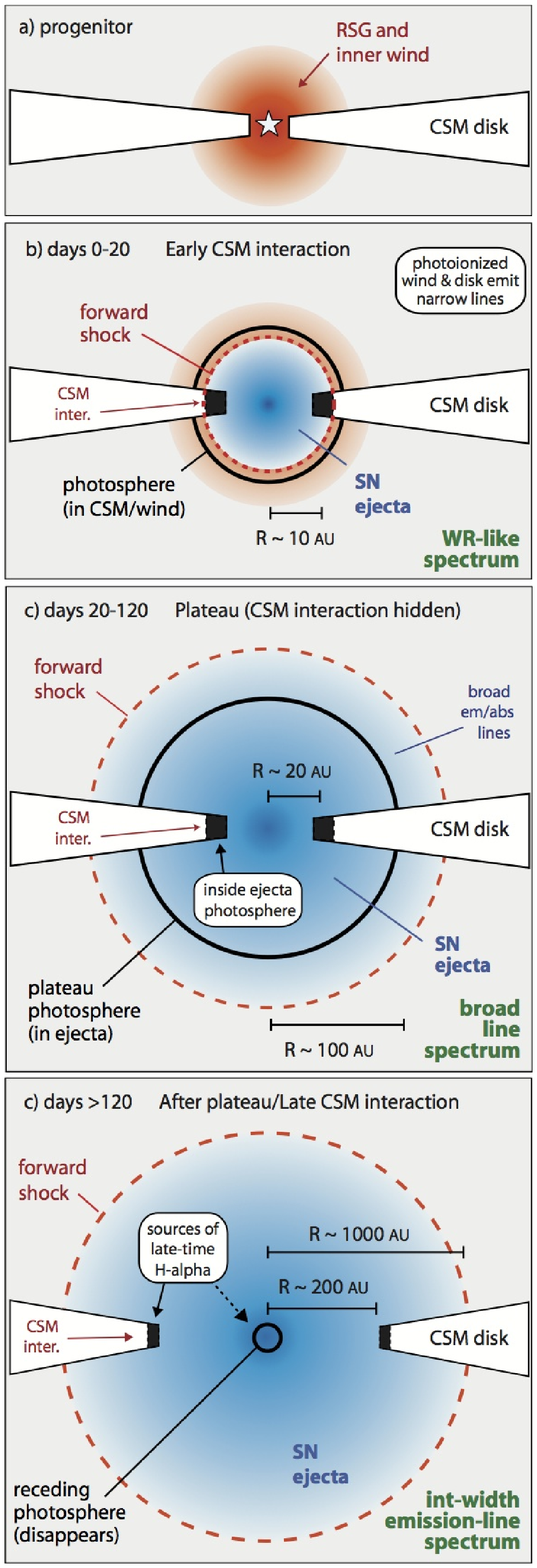}
\caption{Sketch of the more complicated interaction when the CSM is
  highly asymmetric; in this example a disk-like distribution
  surrounds the progenitor. (a) The initial pre-explosion state is a
  progenitor with a wind, as well as a dense disk at radii of $\sim$10
  AU.  (b) Immediately after explosion, narrow lines may arise either
  from a photoionized dense wind or from the disk.  Strong CSM
  interaction with the disk occurs immediately and enhances the early
  luminosity, but the emergent spectrum may be dominated by narrow
  lines with Lorentzian wings.  (c) CSM interaction with the disk has
  slowed the forward shock in the equatorial plane, but the SN ejecta
  expand relatively unimpeded in the less dense polar regions.  After
  a few days, the normal SN ejecta photosphere may expand so much that
  it completely envelopes the CSM interaction occurring in the disk.
  The enveloped CSM interaction can now heat the optically thick
  ejecta from the inside and contribute significantly to the total
  visible luminosity, even if no narrow lines are seen, and may cause
  the photosphere to be asymmetric.  (d) At late times, the SN ejecta
  photosphere recedes.  At this time, the CSM interaction in the disk
  is exposed again (now with larger radial velocities because it has
  been accelerated), and intermediate-width and possibly
  doubled-peaked lines may be seen from the ongoing interaction. This
  sketch was originally invoked to explain the observed evolution of
  PTF11iqb (Smith et al.\ 2015).}
\label{fig:11iqb}     
\end{figure}

This sort of analysis is oversimplified, but it is sufficient to give
us a ballpark idea of the sorts of mass-loss rates that are required
to produce a Type~IIn supernova.  In order to make the CSM-interaction
luminosity high enough that it can compete with the photosphere of a
typical SN (say 3$\times$10$^8$ $L_{\odot}$, or $-$16.5 mag), and with
representative values of $V_{\rm CSM}$ = 100 km s$^{-1}$ and $V_{\rm
  CDS}$ = 2500 km s$^{-1}$, we require a huge mass-loss rate of order
0.01 $M_{\odot}$ yr$^{-1}$.  SNe~IIn tend to be more luminous than
$-$16.5 mag, or even super-luminous, and so the implied mass-loss
rates get very big (although remember that $V_{\rm CDS}$ is probably
faster than 2500 km s$^{-1}$ in the early optically thick phases near
peak).  This mass-loss rate is well beyond the realm of normal stellar
winds (Smith 2014).  We return to the implications for progenitors
later.

For a constant mass-loss rate in the progenitor wind (and hence, a
single value of $w$), $L$ will usually drop with time as the forward
shock decelerates.  Equation (1) does a pretty good job of describing
the visible continuum luminosity as long as the material is optically
thick enough to reprocess most or all of the X-ray and UV luminosity
into visible continuum radiation.  It doesn't work so well at very
early times during the rise or at peak when diffusion is important,
and it doesn't work at later times when the X-rays begin to leak out
and the H$\alpha$ line emission becomes very strong.  At these phases,
Equation (2) will underestimate the required progenitor mass-loss rate
if $L$ is derived from the visible light curve. At later times, one
needs to adopt a very uncertain efficiency factor in order to derive
physical parameters from observations.  The efficiency of converting
kinetic energy into visible light can vary widely depending on CSM
density (van Marle et al.\ 2010), CSM composition (H-free gas tends to
be more transparent), and geometry, as discussed next.

\section{A (Probably) More Realistic Physical Picture}
\label{sec:3}

Figure~\ref{fig:06tf} is a good place to begin when picturing the
physics of CSM interaction, but it may be something of a fantasy.  One
exceedingly obvious result from 20 years of {\it Hubble Space
  Telescope} imaging of circumstellar shells is that they are often
not spherically symmetric, and this is especially true of massive
stars and eruptive mass loss.  Bipolar shells and equatorial disks or
rings are common around evolved massive stars.

This has an impact on our interpretation of observed signatures of CSM
interaction, because when spherical symmetry is broken, it can affect
the qualitative appearance of the spectrum, and it can change line
profile shapes.  More importantly, it can dramatically change the
inferred total energy and mass budgets, and sometimes it can hide
evidence of CSM interaction altogether.  If one assumes spherical
symmetry in an interacting SN that is actually not spherically
symmetric, one can be led to astoundingly wrong conclusions.  In this
case, the ``simplest'' assumptions may be simply inappropriate.

Figure~\ref{fig:11iqb} shows an example of how the scenario in
Figure~\ref{fig:06tf} might be modified if a SN is surrounded by an
equatorial disk-like distribution of CSM as opposed to a spherical
shell (this is borrowed from Smith et al.\ 2015, where much of this
topic is discussed in more detail).  A key point is that if the dense
CSM is disk-like, then strong CSM interaction only occurs in a small
fraction of the total solid angle involving the equatorial regions
(see Mauerhan et al.\ 2014; Smith et al.\ 2014).  The vast majority of
the SN ejecta (say, 90\% if the disk only intercepts 10\% of the solid
angle seen by the SN) can expand unimpeded in the polar directions.
Thus, as far as the majority of the SN ejecta are concerned, their
kinetic energy is not tapped by CSM interaction, and they would emit a
normal broad-lined spectrum.  For the unlucky 10\% of the SN ejecta in
the equator, their kinetic energy is converted into heat and radiation
by the shock.  In this way, CSM interaction with a disk can create SNe
that are overluminous compared to normal SNe II-P, for example, but
not nearly as luminous as the class of super-luminous SNe.  This may
be applicable to many SNe~IIn.

{\bf Enveloped, Swallowed, and Hidden CSM Interaction.}  There is a
nonintuitive but very interesting result that can occur from
interaction with asymmetric CSM, which is that {\it the resulting SN
  might not show any narrow lines even if much of its luminosity comes
  from strong CSM interaction.}  This can occur in the following way:
Imagine that a SN progenitor is surrounded by a dense CSM disk located
at a radius 10-20 AU.  The initial shock interaction might produce
strong narrow lines in the first few days.  In the disk, the fast SN
ejecta are decelerated, creating a strong shock that produces a
moderately high luminosity and bright narrow-line emission from this
CSM interaction.  However, the remaining 90\% of the SN ejecta expand
unimpeded as if there was no CSM interaction.  The normal SN ejecta
photosphere quickly expands to a radius of 10s of AU in a few days.
At such time, the optically thick SN ejecta can completely envelope
the equatorial CSM interaction, hiding some or all of the narrow and
intermediate-width lines that arise due to CSM interaction.  The extra
radiated energy generated by the equatorial CSM interaction is now
unable to escape freely because it has been swallowed by the opaque
fast ejecta, and this radiative energy is instead deposited in the
interior of the optically thick SN ejecta and must diffuse out.  It
will therefore add extra heat to the outer envelope, perhaps mimicking
extra deposition of radioactive decay energy.  The asymmetric
injection of heat in this way will likely make the photosphere
asymmetric.  Viewed from the outside at a time of roughly a month
after explosion, one might then observe a broad-lined SN photosphere
with enhanced luminosity and perhaps a longer than normal duration,
unusual line profiles, or significant polarization.  At this stage, it
would be difficult to tell the difference between some small extra
amount of radioactive heating or buried CSM interaction.

All hope is not lost, though, because an observer might then be able
to deduce that CSM interaction was contributing to the main light
curve all along by watching until late times.  After the SN
recombination photosphere recedes, the enveloped CSM interaction
region is exposed once again.  When this happens, the dense CSM disk
that has been swept up into a cold dense shell (or cold dense torus)
and accelerated to a few thousand km s$^{-1}$ may now be seen again
via strong intermediate-width line emission.  Prominent examples of
this are SN~1998S, PTF~11iqb, SN~2009ip, and even the SN~IIb 1993J
(Leonard et al.\ 2000; Pozzo et al.\ 2004; Smith et al.\ 2014, 2015;
Mauerhan et al.\ 2014; Matheson et al.\ 2000).  This line emission can
be much stronger than the nebular emission from the optically thin
inner ejecta heated by the dwindling radioactive decay.  Moreover, the
ongoing CSM interaction can emit strong X-rays that propagate back
into the SN ejecta, potentially changing the emergent nebular
spectrum.

Perhaps the most important thing to realize is that we now have two
different physical explanations for the wide diversity in peak
luminosity of SNe~IIn, even if all explosions have the same 10$^{51}$
ergs of kinetic energy.  The range of luminosity from the strongest
CSM interaction in super-luminous SNe down to cases with minimal extra
luminosity provided by CSM interaction can be explained either by
ramping down the density of the progenitor's wind, or by increasing
the degree of asymmetry in dense CSM.  The strongest clues that
significant CSM asymmetry is present are from: (1) spectropolarimetry
(e.g., Mauerhan et al.\ 2014), (2) asymmetric line profile shapes
(although some asymmetry can also be caused by dust or occultation by
the SN photosphere itself), or (3) the time evolution of velocities
(for example, seeing very fast speeds in the SN ejecta that persist
after the strongest CSM interaction has subsided; Smith et al.\ 2014).
The last point confirms unequivocally that not all the fast SN ejecta
participated in the CSM interaction.  There may also be some special
observed signatures of asymmetry at certain viewing angles (edge-on,
for example) and specific times.  If one does recognize such
signatures of asymmetry, one must realize that the progenitor
mass-loss rate inferred from equation (2) as well as the SN explosion
energy are not just lower limits, but likely underestimates by a
factor of $\sim$10.  Of course, one can also invoke differences in
explosion energy, explosion mechanism, SN ejecta mass, and radioactive
yield to also contribute to the observed diversity.

\section{Observed Subtypes}
\label{sec:4}
The main physical parameters that determine the observed properties of
a typical core-collapse SN are the mass of ejecta, kinetic energy of
the explosion (as well as the mass of synthesized radioactive
material), and the composition and structure of the star's envelope at
the time of explosion.  This leads to the wide diversity in observed
types of normal ejecta-dominated core-collapse SNe and thermonuclear
SNe (II-P, II-L, IIb, Ib, Ic, Ia).

For interacting SNe, we have all these same free parameters of the
underlying explosion, as well as the possibility of SNe with low or no
radioactive yield and a wider potential range of explosion energy ---
but to these we must also add the variable parameters associated with
the CSM into which the SN ejecta crash: CSM mass and/or mass-loss
rate, radial distribution (speed and ejection time before explosion),
CSM composition, and CSM geometry.  Given these parameters, it may not
be surprising that the class of interacting SNe is extremely diverse,
and observations are continually uncovering new or apparently unique
characteristics.  Moreover, a SN can change type depending on when it
is seen.  An object that looks like a SN~IIn in the first few days can
morph into a normal SN~II-L or IIb if it undergoes
enveloped/asymmetric CSM interaction, and may then return to being a
Type~IIn at late times, as discussed in the previous section.

There are, however, some emerging trends among interacting SNe.  The
list below attempts to capture some of these emerging subtypes among
interacting SNe that appear to share some common and distinct traits.
The reader should be advised, however, that this is still a rapidly
developing field, so this is neither a definitive nor a complete list,
and moreover, there are observed events that seem to skirt boundaries
or overlap fully between different subtypes.  Further subdivisions are
likely to be clarified with time.  The list below includes a
descriptive name and some prototypical or representative observed
examples that are often mentioned (again, not a complete list).

\runinhead{Superluminous IIn; compact shell (SN~2006gy).}  SN~2006gy
was the first superluminous SN to be discovered, and it remains
extremely unusual even though several other SLSNe~IIn have since been
found (see Smith et al.\ 2007, 2010a; Ofek et al.\ 2007; Woosley et
al.\ 2007). It had a slow rise to peak ($\sim$70 days) and faded
within another 150 days or so, which is unlike the other SLSNe IIn
that seem to fade very slowly and steadily from peak.  Also unusual
was that it had strong intermediate-width P Cygni absorption features
in its spectra, strong line-blanketing absorption in the blue, and
narrow P Cygni features from the CSM.  It is thought to have arisen
from a relatively compact and opaque CSM shell, where CSM interaction
mostly subsided within a year. In this case, the CSM is thought to
have arisen from a single eruption that ejected $\sim$20 $M_{\odot}$
about 8 years prior to core collapse (Smith et al.\ 2010a).  SN~2006gy
is one of the best observed SNe~IIn and is often discussed, but
readers should be aware that it is not at all typical.

\runinhead{Superluminous IIn; extended shell (SN~2006tf, SN~2010jl,
  SN~2003ma).}  These are basically the superluminous version of the
SN~1988Z-like subclass, discussed next.  Unlike SN~2006gy, they show
strong, smooth blue continua without much line blanketing, and little
or no P Cygni absorption in their intermediate-width lines.  The also
tend to fade slowly and steadily, in some cases remaining bright for
several years as the shock runs through an extended dense shell (e.g.,
Smith et al.\ 2008; Rest et al.\ 2011; Fransson et al.\ 2014).

\runinhead{Enduring IIn (SN~1988Z, SN~2005ip).}  These SNe~IIn show a
smooth blue continuum superposed with strong narrow and
intermediate-width H lines, and in some cases even broad components
(Chugai \& Danziger 1994; Aretxaga et al.\ 1999; Smith et al.\ 2009).
Some cases, such as SN~2005ip, have strong narrow coronal emission
lines (Smith et al.\ 2009), implying photoionization of dense clumpy
CSM by X-rays generated in the shock.  They tend to be more luminous
than SNe II-P, but not as bright as SLSNe.  These objects tend to fade
very slowly and show signs of strong CSM interaction for years or even
decades after discovery; SN~1988Z is {\it still} going strong.  One
can picture these as a stretched-out version of the SLSNe~IIn, in the
sense that they have somewhat lower luminosities at peak, but they
last longer, eventually sweeping through similar amounts of total CSM
mass (of order several to 20 $M_{\odot}$).  While SLSNe have CSM that
is so dense that it requires eruptive LBV-like mass loss in the years
or decades before explosion, these enduring SNe~IIn like SN~1988Z and
SN~2005ip can potentially be explained by extreme RSG winds blowing
for centuries before core colllapse.  If there is such a thing as a
``standard'' SN~IIn, this is probably what most IIn-enthusiasts have
in mind.

\runinhead{Transitional IIn (SN~1998S, PTF11iqb, SN~2013cu).}  This
class of SNe~IIn only shows fleeting signatures of CSM interaction
that disappear quickly, and may be entirely missed if the SN isn't
discovered early after explosion.  SN~1998S was one of the nearest and
best studied SNe~IIn that helped shape our understanding of SNe~IIn,
and so it has often been referred to as ``prototypical'' -- but it
really isn't. SN~1998S wasn't very luminous and its spectral
signatures of CSM interaction disappeared pretty quickly,
transitioning into a broad-lined ejecta-dominated photosphere within a
couple weeks (Leonard et al.\ 2000; Shivvers et al.\ 2015).  This
indicated that the total mass of CSM was actually quite modest (of
order 0.1 $M_{\odot}$ or so), substantially different from the types
noted above.  If it had not been caught so early, SN~1998S might not
have been classified as a Type~IIn at all.  An older example of this
was SN~1983K (Nielmela et al.\ 1985).  More recently, a few other
related objects have been found (SN~2013cu \& PTF11iqb; Gal-Yam et
al.\ 2014; Smith et al.\ 2015), which only showed a Type~IIn spectrum
for the first few days after discovery (and these cases were thought
to have been discovered very early; within 1 day or so of explosion).
PTF11iqb and SN~2013cu then morphed into broad lined SNe with spectra
similar to SNe II-L or IIb, respectively.  At late times, PTF11iqb
showed strong CSM interaction again, very similar to SN~1998S and the
Type IIb SN~1993J. We do not know how common these are.  While SNe~IIn
represent roughly 8-9\% of core collapse SNe (Smith et al.\ 2011),
this statistic does not include SNe that only exhibit Type~IIn
characterstics for a brief window of time and then morph into other
types after a few days; thus, early strong CSM interaction with dense
inner winds might be fairly common among the larger population of
ccSNe.  The corresponding (potentially very important) implication is
that many core collapse SNe may suffer a brief pulse of episodic mass
loss shortly before core collapse.  The reason for this is not yet
known, but is probably related to the final nuclear burning stages and
may have important implications for models of core collapse.

\runinhead{Late-time interaction in otherwise normal SNe (SN 1993J,
  SN~1986J).} The overlap with the previous subclass is probably
considerable, but it is worth mentioning that some otherwise normal
SNe (perhaps appearing normal simply because we missed the early IIn
signatures in the first day or so) show particularly strong CSM
interaction at late times.  Some of the more common cases are SNe IIb
and SNe II-L, for which there are well-studied and famous nearby
examples like SN~1993J and SN~1986J (Matheson et al.\ 2000;
Milisavljevic \& Fesen 2013).  For these, the transition from a SN
into a SN remnant is somewhat blurry, and some of these are well known
nearby aging SNe.  Because this interaction is most apparent when the
SNe have faded after the first year, this phenomenon can only be
studied in nearby galaxies.  These may be caused by strong RSG winds,
or by equatorial CSM deposited by binary interaction in objects like
SNe~IIb that are more clearly associated with the end products of
binary evolution.  The reason why this CSM still resides close to the
star at the time of death is still unclear, and (as for the previous
subclass) may hint at some rapid and dramatic changes in the final
nuclear burning sequences in a wider fraction of SN progenitors than
just standard SNe~IIn.

\runinhead{Delayed onset, slow rise, multi-peaked (SN~2009ip,
  SN~2010mc, SN~2008iy, SN~1961V, SN~2014C).}  In contrast to the
previous subclass, some SNe IIn show little or no signs of CSM
interaction at first, but then rise to the peak of CSM interaction
after a delay of months or years.  This delay is presumably due to the
time it takes for the fastest SN ejecta to catch up to a CSM shell
that was ejected a year or more before core collapse.  The underlying
SN ejecta photosphere could be relatively faint at first if the
progenitor was a more compact BSG, like SN~1987A, or an LBV; the
faintness of the initial SN might cause the initial transient to be
missed altogether, or mistaken as a pre-SN eruption rather than the
actual SN, since it might precede the delayed onset of peak luminosity
that actually arises when the ejecta overtake the CSM.  It must be
admitted, however, that from the time of peak onward, these tend to
appear as relatively normal SNe IIn; as such, the distinction between
this ``delayed onset'' subclass and other SNe IIn might be artificial,
and only distinguishable in cases with fortuitous pre-discovery or
pre-peak data.  The amount of delay in the onset of CSM interaction
bears important physical information about the elapsed time between
the pre-SN eruption and core collapse.  In the case of SN~2009ip, the
delay of $\sim$40 days made sense, as fast SN ejecta caught up to CSM
moving at 10\% the speed, associated with eruptions that were actually
observed a year or so before the SN (Mauerhan et al.\ 2013a, 2014;
Smith et al.\ 2014).  SN~1961V had a light curve consistent with a
normal SN~II-P for the first $\sim$100 days, followed by a brighter
peak (Smith et al.\ 2011b). In SN~2014C, the delayed onset had
different chemical properties, where a H-poor stripped-envelope SN
crashed into a H-rich shell after a year (Milisavljevic et al.\ 2015).
In some cases, however, the delayed onset is rather extreme and the
CSM interaction strong; SN~2008iy appeared to be a relatively normal
core collapse SN in terms of luminosity, which rose to very luminous
peak as a Type IIn after 400 days (Miller et al.\ 2010).  SN~1987A
might be thought of as an even more extended version of this, where
the onset of CSM interaction was delayed for 10 years as the SN ejecta
caught up to CSM ejected $\sim$10$^4$ yr earlier.  We don't know what
fraction of normal SNe have a delayed onset of CSM interaction, since
most SNe are not monitored continuously with large telescopes for
years or decades after they fade.

\runinhead{Type IIn-P (SN~1994W, SN~2011ht, the Crab).}  This is a
distinct subclass of SNe~IIn that exhibit IIn spectra throughout their
evolution, but have light curves with a well-defined plateau drop
(Mauerhan et al.\ 2013b).  They are not to be confused with
``transitional'' SNe IIn/II-P that may have narrow lines at early
times and evolve into an otherwise normal SN~II-P.  In SNe~IIn-P, the
narrow H lines with strong narrow P Cyg profiles are seen for the
duration of their bright phase.  The drop from the plateau is quite
extreme (several magnitudes) and the late-time luminosities suggest
low yields of $^{56}$Ni.  These may be the result of $\sim$10$^{50}$
erg electron capture SN explosions with strong CSM interaction
(Mauerhan et al.\ 2013b; Smith 2013).  It has been proposed that the
Crab Nebula may be the remnant of this type of event, although this
remains uncertain (Smith 2013).


\runinhead{SN~IIn impostors (LBVs, SN~2008S-like, etc.).}  These
transients have narrow H lines in their spectra similar to SNe~IIn,
but are subluminous (fainter than $M_R \simeq -$15.5 mag), and have
slower velocities than normal SNe~IIn (see Smith et al.\ 2011b and
references therein).  One may argue that the luminosity cut is
arbitrary, but the narrow lines without any broad wings do seem
distinct from other SNe~IIn (most of the time).  Some are certainly
non-terminal transients akin to LBVs (because they repeat or the star
survives), but some cases are not so clear.  Although they may not be
SNe, they are included here because we're not sure yet --- some
objects thrown into this bin could be underluminous because they are
interacting transients that arise from ``failed'' SNe (fallback to a
black hole), pulsational pair instability SNe, ecSNe, SNe from compact
BSG progenitors with low radioactive yield, mergers with compact
object companions, or other terminal events.

\runinhead{Type Ia/IIn or Ia-CSM (SN~2002ic, SN~2005gj, PTF11kx).}
These are transients that show spectral features indicative of an
underlying SN Ia ejecta photosphere, but with strong superposed narrow
H lines and additional continuum luminosity (Silverman et al.\ 2013).
Cases with stronger CSM interaction tend to obscure the Type Ia
signatures, leading to ongoing controversy about their potential core
collapse or thermonuclear nature (Benetti et al.\ 2006).  Cases with
weaker CSM interaction (PTF11kx) reveal the Type Ia signatures more
clearly, allowing them to be identified unambiguously as thermonuclear
events.  The relatively rare evolutionary circumstance that leads a
thermonuclear SN to have a large mass of H-rich CSM is still unclear.

\runinhead{Type Ibn (SN~2006jc).} These are similar to SNe~IIn, except
that H lines are weak or absent, so that narrow or intermediate-width
He~{\sc i} lines dominate the spectrum (H$\alpha$ is weaker than
He~{\sc i} $\lambda$5876).  Without the assistance of H opacity, these
objects tend to fade more quickly than most SNe~IIn. The class is
quite diverse and includes a range of H line strengths and CSM speeds,
including transitional IIn/Ibn cases (SN~2011hw) where H and He~{\sc
  i} lines have similar strength (Smith et al.\ 2012; Pastorello et
al.\ 2008, 2015). It has been hypothesized that the likely progenitors
may be Wolf-Rayet stars or LBVs in transition to a WR-like state
(Pastorello et al.\ 2007; Foley et al.\ 2007; Smith et al.\ 2012).

\runinhead{Type Icn (hypothetical).}  A discovery of this class has
not yet been reported (as far as the author is aware), but as
transient searches continue, there may be cases where a SN interacts
with a dense shell of CSM that is both H and He depleted, yielding
narrow or intermediate-width lines of CNO elements, for example.  With
sufficient creativity, one can imagine stellar evolution pathways that
might create this; such SNe are clearly rare, and if they never turn
up, their absence will provide interesting physical constraints for
some binary models.

\section{Dust Formation in CSM interaction}

The formation of dust by SNe may be essential to explain the amount of
dust inferred in high redshift galaxies.  SNe with strong CSM
interaction provide an avenue for dust formation that is different
from normal SNe, and possibly much more efficient.  In normal SNe,
dust forms in the freely expanding ejecta where there is a competition
between cooling to low enough temperatures while the ejecta are also
rapidly expanding and achieving lower and lower densities.  Even if
dust forms efficiently, this ejecta dust might get destroyed when it
crosses the reverse shock.  In interacting SNe, by contrast, evidence
suggests that dust can form very rapidly in the extremely dense,
post-shock cooling layers (zones 2 and 3 in Figure~\ref{fig:06tf}).
Moreover, this dust is already behind the shock and may therefore
stand a better chance of surviving and contributing to the ISM dust
budget.

The first well-studied case of post shock dust formation was in
SN~2006jc, which was a Type Ibn.  The classic signs of dust formation
were seen starting at only 50 days post-dicovery, in an increase in
the rate of fading, excess IR emission, and an increasing blueshifted
asymmetry in emission-line profiles (Smith et al.\ 2008b).  The last
point strongly favored post-shock dust formation, since this was seen
in the intermediate-width lines that were emitted from the post-shock
zones.  The Type Ibn event may have been the result of a Type~Ic
explosion crashing into a He-rich shell.  In this case, one might
think that the C-rich ejecta were important in assisting the efficient
formation of dust, as is seen for the post-shock dust formation in
colliding-wind WC+O binaries (Gehrz \& Hackwell 1974; Williams et al.\
1990).  However, similar evidence for this same mode of post-shock
dust formation has also been seen in several Type IIn events, such as
SN~2005ip, SN~1998S, SN~2006tf, SN~2010jl, and others (Smith et al.\
2008a, 2009, 2012; Fox et al.\ 2009; Gall et al.\ 2014; Pozzo et al.\
2004).  It may therefore be the case that enhanced post shock dust
formation is a common outcome of strong CSM interaction, where
efficient post-shock cooling causes the forward shock to collapse and
become very dense.

\begin{figure}[h]
  \sidecaption
\includegraphics[scale=.5]{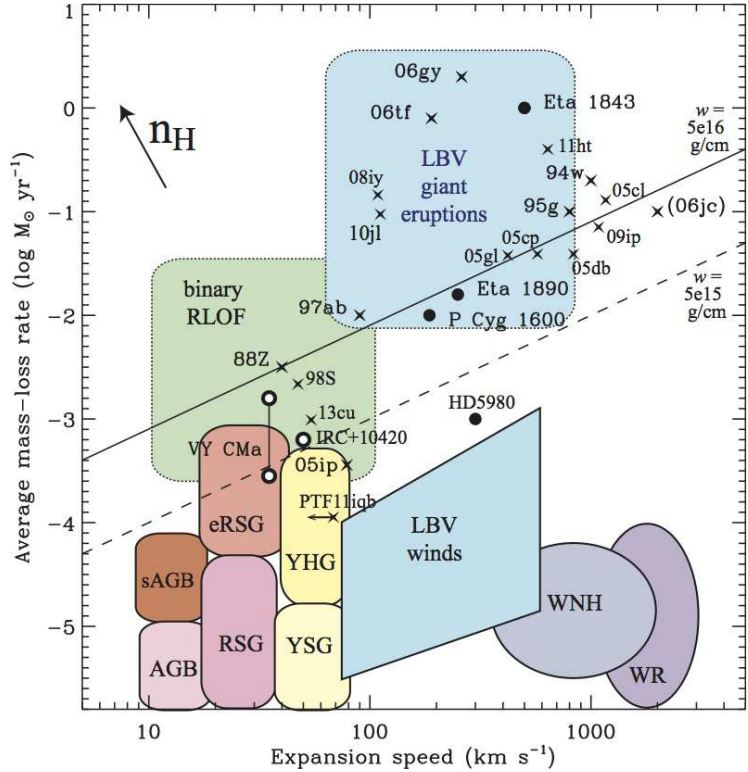}
\caption{Plot of mass-loss rate as a function of wind velocity,
  comparing values for interacting SNe to those of known types of
  stars.  The solid colored regions correspond to values for various
  types of evolved massive stars taken from the review about mass-loss
  by Smith (2014), corresponding to AGB and super-AGB stars, red
  supergiants (RSGs) and extreme RSGS (eRSG), yellow supergiants
  (YSG), yellow hypergiants (YHG), LBV winds and LBV giant eruptions,
  binary Roche-lobe overflow (RLOF), luminous WN stars with hydrogen
  (WNH), and H-free WN and WC Wolf-Rayet (WR) stars.  A few individual
  stars with well-determined very high mass-loss rates are shown with
  circles (VY CMa, IRC+10420, $\eta$ Car's eruptions, and P Cyg's
  eruption).  Also shown with X's are some representative examples of
  SNe~IIn (and one SN~Ibn) that have observational estimates of the
  pre-shock CSM speed from the narrow emission component as measured
  in moderately high resolution spectra as well as estimates of the
  progenitor mass-loss required, taken from the literature.  The
  diagonal lines are wind density parameters ($w$=$\dot{M}/V_{\rm
    CSM}$) of 5$\times$10$^{16}$ and 5$\times$10$^{15}$ g cm$^{-1}$,
  which are typically the lowest wind densities required to make a
  SN~IIn.  Values are taken from the literature; this figure is from a
  paper in prep.\ by the author.  Note that in some cases, an
  observationally derived value for the mass of the CSM has been
  converted to a mass-loss rate with a rough estimate of the time
  elapsed since ejection.}
\label{fig:mdot}     
\end{figure}

\section{Progenitors and Pre-SN Mass Loss}
Basic considerations about powering SNe~IIn with CSM interaction,
noted above, argue that extremely strong pre-SN mass loss is
required. What sorts of progenitor stars can give rise to this dense
CSM?

\subsection{CSM Properties}

The first thing to do is to look around us in the nearby universe and
ask what sort of observed classes of stars might fit the bill.
Figure~\ref{fig:mdot} makes this comparison by plotting the diversity
in wind density of interacting SNe, deduced from the pre-shock wind
(or shell) expansion speed and the inferred mass-loss rate, as
compared to expected mass-loss from known types of stars.  The shaded
and colored regions in this figure show rough parameters for different
classes of evolved massive stars that are potential SN progenitors
with strong winds.  These are taken from the recent review by Smith
(2014); see the caption for definitions.  In this plot, a given wind
density parameter (required for a particular value of a CSM
interaction luminosity via Equation 1) has a diagonal line increasing
to the upper right.  The solid line shows a typical value for a
moderate-luminosity SN IIn, and the dashed line is a typical lower
bound for wind densities required to make a SN IIn (although an object
can be slightly below this line and still make a Type IIn spectrum if
the wind is clumpy or asymmetric).

We can see immediately that the normal classes of evolved stars with
strong but relatively steady winds (WR stars, LBV winds, normal YSGs
and RSGs, AGB stars) do not match up to the wind density required for
SNe~IIn.  These could potentially produce SNe that have strong X-ray
or radio emission from CSM interaction, but they are not dense enough
to slow the forward shock, and to cause the forward shock to cool
radiatively and collapse into the cool dense shell that is required
for intermediate-width H$\alpha$.  The only observed classes of stars
that have the high wind densities required are the giant eruptions of
LBVs and the most extreme cool hypergiants (the slower winds of
extreme RSGs and YHG make their winds have density comparable to the
shells in LBV giant eruptions, even though the mass-loss rates are
lower).  Both classes of stars are not steady winds, but rather, they
are dominated by relatively short-lived phases of eruptive or episodic
mass loss, or extreme dense and clumpy winds.  In the case of LBVs,
these are eruptions that last a few months to a decade.  For extreme
RSGs and YHGs, these are phases of enhanced mass loss that may persist
for centuries to a few thousand years.

From the point of view of their mass loss, LBVs (and to a lesser
extent the most extreme cool hypergiants) are good candidates for the
progenitors of SNe~IIn.  This does not, however, mean that these stars
are necessarily all poised to explode in the next couple years, nor
does it mean that they are the only possible IIn progenitors.  The
caveat is that the progenitors of SNe~IIn undergo strong eruptive or
episodic mass loss immediately before core collapse, and may therefore
experience significant changes in their internal structure.  The stars
may have looked very different in the time period before their pre-SN
eruptions began.  Hence, these stars in their immediate pre-SN state
may not actually exist among nearby populations of stars in the Milky
Way and Magellanic Clouds right now.

Also note that this is a mass-loss {\it rate}, not total mass, and
note that these are usually lower limits to the required mass-loss
rate (the high rate makes high optical depths, and hence, makes an
observational determination difficult).  In cases where we have
estimates of the total mass ejected, one must divide by an assumed
timescale to derive a mass-loss rate; for SNe~IIn, this timescale is
inferred from the relative velocities of the CSM and CDS.  Since we
are essentially plotting a wind density, it is important to recognize
that two different SNe with the same inferred progenitor mass-loss
rate might have had that high mass loss lasting for very different
amounts of time, and hence, may have very different amounts of total
mass ejected shortly before the SN.  In some cases, the total mass can
be more constrainting than the rate.  For example, the total mass of
$\sim$20 $M_{\odot}$ required for some super-luminous SNe IIn rules
out any progenitor stars with initial masses below about 30-40
$M_{\odot}$, because we need a comparable mass of SN ejecta to provide
enough momentum for a long-lasting interaction phase (not to mention
mass loss due to winds throughout the life of the star).  Also, there
are cases like SN~2005ip where the inferred mass-loss rate is on the
low end, but this SN had remarkably long-lasting CSM interaction that
suggests a total mass of order 15 $M_{\odot}$ of H-rich material
(e.g., Smith et al.\ 2009; Stritzinger et al.\ 2012), which rules out
lower-mass RSGs and super-AGB stars for its progenitor.

\begin{svgraybox}
  Warning 3.  One must be somewhat cautious in drawing conclusions
  about the progenitor based on only the observed pre-shock CSM speed.
  We generally think of the outflow speed being proportional to the
  star's escape velocity, so we expect slower outflows from RSGs and
  YHGs, somewhat faster speeds for LBVs and BSGs (few 10$^2$ km
  s$^{-1}$), and very fast outflows for WR stars (few 10$^3$ km
  s$^{-1}$).  This is a good guide if the outflow is a relatively
  steady radiation-driven wind.  However, remember that the observed
  CSM may be the result of eruptive or explosive mass loss driven by a
  shock wave.  If so, it is possible to get a relatively fast outflow
  even from a bloated RSG; eruption speeds need not match steady wind
  speeds for various types of stars in Figure 3.  Moreover, the high
  luminosity of the SN itself can potentially accelerate the pre-shock
  CSM (essentially a radiatively driven wind from a temporarily much
  more luminous star), which could also lead to a faster pre-shock CSM
  speed than expected for a certain type of star, or multiple velocity
  components in the CSM.
\end{svgraybox}

\subsection{Direct Detections}

The types of inferences described above are all still based on rather
indirect and circumstantial evidence.  A potentially more direct way
to link SNe to their progenitors is to detect the progenitor star in
pre-explosion imaging data, usually with the {\it Hubble Space
  Telescope}, although ground-based imaging has been used for some
very nearby cases (see review by Smartt 2009).  This method has been
used successfully for some other types of SNe, especially SNe II-P,
IIb, and one particularly famous II-pec event that occurred 20 years
ago.  The progenitor identification can be confirmed after the SN
fades, to verify that the progenitor star is gone (as opposed to being
a chance alignment, a host cluster, or a companion star).

For interacting SNe, however, the interpretation of pre-SN direct
detections can be a little tricky.  First, the ``direct'' detection of
the progenitor might actually be a direct detection of a pre-SN
eruption and not the quiescent star.  This source might indeed fade
after the SN, but it is hard to tell if the star is really dead, or if
the eruption has just subsided and the star returned to its quiescent
state.  Inferring an initial mass by comparison with stellar evolution
tracks is also complicated if the progenitor might be in an outburst
rather than in its quiescent state (we don't have much choice here and
are lucky if there is even one {\it HST} image in the archive, but one
just needs to be aware of the caveat).  A second complication is that
some SNe IIn have persistent CSM interaction for years or decades
after the SN, and so one might need to wait a very long time before
the SN has faded enough to be fainter than the progenitor.  Third, a
faint progenitor or even a faint upper limit to a progenitor is very
inconclusive in terms of its implication for the mass of the star.
This is because with interacting SNe, there is, by definition, a large
mass of CSM.  Thus, we certainly expect some cases where the
progenitor was surrounded by a massive dust shell that should obscure
the progenitor star's visible light output.  This makes it difficult
to place a progenitor detection (or upper limit) on an HR Diagram and
infer an initial mass if one has only an optical filter.

Despite this ambiguity, there have been a few lucky and important
cases that guide our intuition about the progenitors of SNe~IIn and
SNe~Ibn.

{\bf SN~2005gl:} This was a SN~IIn where a very luminous progenitor
consistent with an LBV star like P Cygni was detected in {\it HST}
imaging and the SN had an implied mass-loss rate of 0.01 $M_{\odot}$
yr$^{-1}$ (Gal-Yam et al. 2007), and this source then faded after the
SN (Gal-Yam \& Leonard 2009).  Moreover, the pre-shock CSM speed of
420 km s$^{-1}$ inferred from narrow H lines was suggestive of the
fast outflow from an LBV (Smith et al. 2010).  As noted above, there
is some ambiguity as to whether the pre-SN detection was a massive
quiescent LBV-like star or a pre-SN eruption caught by the {\it HST}
image.  In either case an eruptive LBV-like star is likely, although
the implied initial mass may be different.  Based on its spectral
evolution, SN~2005gl fits into the subclass of ``transitional'' SNe
IIn like SN~1998S and PTF11iqb, which show a Type IIn spectrum at
first, but later show broad P Cygni lines indicative of a SN
photosphere, so this can be taken as an argument that it was most
likely a core-collapse SN event.

{\bf SN~1961V:} Long considered an extreme LBV or SN impostor event,
recent arguments favor an actual core-collapse SN for the 1961
transient (Kochanek et al.\ 2011; Smith et al. 2011).  If this was a
SN, then it has one of the best documented progenitor detections and
progenitor variability among SNe, and it holds the record for the most
massive directly detected SN progenitor.  The pre-1961 variability
suggests a very massive $\sim$100 $M_{\odot}$ blue LBV-like progenitor
that was variable before the SN.  The source at the SN position is now
more than 5.5 mag fainter than this progenitor (a much more dramatic
drop than in the case of SN~2005gl), and there is no IR source with a
luminosity comparable to the progenitor (Kochanek et al.\ 2011), so
the extremely massive star is likely dead.

{\bf SN~2006jc:} An eruption in 2004 was noted as a possible LBV or SN
impostor, and then a SN occurred at the same position 2 years
later. The pre-SN outburst had a peak luminosity similar to SN
impostors (Pastorello et al. 2007). The SN explosion two years later
was of Type Ibn with strong narrow He~{\sc i} emission lines (Foley et
al. 2007, Pastorello et al. 2007). There is no detection of the
quiescent progenitor, but this unusual case implies an LBV-like
eruption that occurred in a WR-like progenitor star that was clearly H
depleted.  The CSM speed was of order 1000 km s$^{-1}$, which is
consistent with WR stars, and faster than typical LBVs.

{\bf SN2011ht:} This belongs to the subclass of SNe IIn-P, which are
thought to arise from lower-energy explosions that may be linked to
electron capture SNe in 8-10 $M_{\odot}$ super-AGB stars (Mauerhan et
al.\ 2013b; Smith 2013).  There is no detection of the quiescent
progenitor, although deep upper limits seem to rule out a luminous,
blue quiescent star (Mauerhan et al.\ 2013b; Roming et al.\ 2012).
However, Fraser et al. (2015) reported the detection of pre-SN
eruption activity in archival data.  This may be an important
demonstration that non-LBVs can have violent pre-SN eruptions as well.

{\bf SN~2010jl:} This was a SLSN IIn with roughly 10 $M_{\odot}$ or
more of CSM.  Smith et al. (2011c) identified a source at the location
of the SN in pre-explosion HST images that suggested either an
extremely massive progenitor star or a very young massive-star
cluster; in either case it seems likely that the progenitor had an
initial mass above 30 $M_{\odot}$.  We are still waiting for this SN
to fade to see if the progenitor candidate was actually the progenitor
or a massive young cluster/association.

{\bf SN~2009ip:} This source was initially discovered as an LBV-like
outburst in 2009 (its namesake) before finally exploding as a much
brighter SN in 2012. A quiescent progenitor star was detected in
archival HST data, indicating a very massive 50-80 $M_{\odot}$
progenitor (Smith et al. 2010b, Foley et al. 2011).  In this case, the
{\it HST} detection may well have been the quiescent progenitor,
rather than an outburst, because much brighter outbursts came later.
It showed slow variability consistent with an S Dor LBV-like episode
(Smith et al. 2010b), followed by a series of brief LBV-like giant
eruptions (Smith et al.  2010b, Mauerhan et al. 2013a, Pastorello et
al. 2013).  Unlike any other object, we also have detailed,
high-quality spectra of the pre-SN eruptions (Smith et al. 2010b,
Foley et al. 2011).  The presumably final SN explosion of SN 2009ip in
2012 would fall into the ``delayed onset'' subclass, since at first
the fainter transient showed very broad lines indicative of a SN
photosphere.  Reaching peak, however, it looked like a normal SN~IIn,
as the fast ejecta crashed into the slow material ejected 1-3 years
earlier (Mauerhan et al. 2013a, Smith et al. 2014). A number of
detailed studies of the bright 2012 transient have now been published,
although there has been some controversy about whether the 2012 event
was a core-collapse SN (Mauerhan et al. 2013a; Ofek et al. 2013;
Prieto et al. 2013; Smith et al. 2014) or some type of extremely
bright nonterminal event (Fraser et al.  2013a, Pastorello et
al. 2013, Margutti et al. 2014). More recently, Smith et al. (2014)
showed that the object continues to fade, and its late-time emission
is consistent with late-time CSM interaction in normal SNe IIn. If SN
2009ip was indeed a SN, it provides a strong case that very massive
stars above 30 $M_{\odot}$ may in fact experience core collapse and
explode, and that LBV-like stars are linked to some SNe~IIn.

\subsection{Links to Progenitor Types}

We will undoubtedly find more examples of direct detections and pre-SN
outbursts in the future.  One must bear in mind, though, that LBVs and
eruptive precursors are relatively easy to detect because they are
brighter than any quiescent stars, so these cases do not rule out
alternatives such as dust-enshrouded RSGs, or faint and hot quiescent
stars.  From various clues described above such as the CSM mass and
mass-loss rate, CSM expansion speed, H-rich composition, and direct
detections of progenitors or environments, we can attempt to link
certain subclasses of interacting SNe to different possible
progenitors.  Some associations are more likely than others.

{\it SLSN IIn (compact and extended shell; SN~2006gy or
  SN~2010jl-like):} Based on the extreme required masses of (10-20
$M_{\odot}$) of H-rich CSM, typically expanding at a few hundred km
s$^{-1}$, it seems very likely that the progenitors of SLSNe IIn are
very massive, eruptive LBV-like stars (see review by Smith 2014).  If
they are not truly LBVs, then they do a very good impersonation.  The
simple fact that very massive stars above $\sim$40 $M_{\odot}$ are
exploding as H-rich SNe is a challenge to understand, since stellar
evolution models predict all those stars to shed their H envelopes at
roughly solar metallicity.  Most of these have huge mass ejections
occurring just a few years or decades before the SN, so the connection
to nearby LBVs --- some of which have shells that are hundreds or
thousands of years old -- is not yet clear.  LBV-like progenitors are
even likely in cases where the progenitor is optically faint, if a
pre-SN eruption has obscured the star with a dust cocoon (as is
likely to be the case, given the consequent CSM interaction).  Since
the progenitors must be very massive stars simply because of the mass
budget, the physical mechanism of pulsational pair instability
eruptions is a viable candidate for the pre-SN outbursts (Woosley et
al.\ 2007).  This is not the case for the remainder of SNe~IIn,
because they are too common (Smith et al.\ 2014).

{\it Normal, enduring SNe~IIn (SN1988Z-like):} Likely progenitor types
are LBVs or extreme cool hypergiants (YHGs/eRSGs), based mostly on the
required mass-loss rates and observed CSM speeds.  For these enduring
cases, we require either strong mass loss in several centuries
preceding the SN, or a large bipolar shell ejected shortly before the
SN with a range of ejection speeds (to accommodate CSM interaction over
a large range of radii, lasting for a long time).  Total CSM masses
that exceed 10 $M_{\odot}$ in some cases and integrated radiated
energies that exceed 10$^{51}$ ergs (over years) point to relatively
massive progenitor stars.  Enhanced late-phase RSG mass loss (Yoon \&
Cantiello 2010) or instabilities in late nuclear burning sequences
(Quataert \& Shiode 2012; Smith \& Arnett 2014) are good candidates
for the episodic mass-loss.

{\it Transitional or fleeting SNe IIn (SN~1998S-like):} Likely
progenitors are RSGs, YHGs, or BSG/LBVs with less extreme pre-SN
bursts of mass loss, confined to a relatively short-duration event
preceding core collapse (i.e. within a few years preceding the SN).
These objects also seem to require highly asymmetric CSM interaction,
to allow the expanding SN photosphere to completely or mostly envelope
the disk of CSM interaction that emits narrow lines (as discussed in
Section 3).  For this reason, there is a strong suspicion that close
binary interaction plays a role in their pre-SN episodic mass loss,
although it must still be linked somehow to the final nuclear burning
sequences (see Smith \& Arnett 2014 regarding the role of close
binaries in this scenario).

{\it SNe~IIn-P (SN~2011ht-like):} The favored scenario for these
events involves an intermediate-mass progenitor in a super-AGB phase
that suffers a sub-energetic (10$^{50}$ erg) electron-capture SN.
This is based on the low $^{56}$Ni yield, the deep absorption features
that imply more-or-less spherical symmetry, and the energy/mass
budgets inferred (Mauerhan et al.\ 2013b; Smith 2013).  However,
observations cannot yet confidently rule out a more massive progenitor
that suffers fallback to a black hole, yielding a smaller ejecta mass
and very low $^{56}$Ni yield (although in the case of SN~2011ht,
progenitor upper limits seem to argue against this interpretation).
This type of SN also fits the bill for SN~1054 and the Crab nebula,
whose abundances and kinetic energy seem to point to an electron
capture SN from an intermediate-mass star (Nomoto et al.\ 1982).  If
these are ecSNe, then the pre-SN episodic mass-loss might be related
to nuclear flashes in the advanced degenerate core buring sequences.

{\it SN IIn impostors}: This group of putatively non-SN transients may
be quite diverse, and it may include transients that have narrow lines
because they are powered largely by CSM interaction (and weaker
explosions than core-collapse SNe) and other transients that have
narrow lines because they have slow winds/outflows.  This may include
LBVs, super AGB stars, binaries with a compact object, stellar
mergers, pre-SN nuclear burning instabilities, failed SNe, or all of
the above.  Any massive supergiant star enshrouded in dust or with
strong binary interaction that suffers an instability is a viable
candidate.  The theoretical mechanisms for this class of outbursts is
still very poorly understood, and there is probably considerable
overlap with pre-SN outbursts that lead to SNe~IIn and Ibn.

{\it SNe Ibn:} The most likely progenitors are massive WR stars that
for unknown reasons undergo pre-SN outbursts.  This is interesting,
because there is no known precedent for an observed eruption in a
H-deficient massive star.

{\it SNe Ia/IIn or Ia-CSM:} If these really are thermonuclear SNe Ia
(some cases seem clear, others are still debated in the literature),
then the exploding progenitors are white dwarfs that have arisen from
initial masses below roughly 8 $M_{\odot}$.  In this case, a single
degenerate system is clearly required to supply the large mass of
H-rich CSM (several $M_{\odot}$ in some cases).  There is, as yet, no
viable explanation for the sudden ejection of a large mass of H (by
the companion) shortly preceding the thermonuclear explosion of the
WD.


\section{Closing comments}

Observations of interacting SNe present one of the most interesting
challenges to our understanding of the end phases of evolution for
massive stars.  What makes these stars explode before they explode?
Computational resources are only beginning to meet the needs of the
complex problem of simulating convection and nuclear burning coupled
with stellar structure and turbulence in these final phases (Arnett \&
Meakin 2011; Meakin \& Arnett 2007).  The fact that $\sim$10\% of
core-collapse events are preceded by some major reorganization of the
stellar structure and energy budget tells us that we have been missing
something important, which might be a key ingredient for understanding
core collapse SNe more generally.  It will be important to try and
understand if this 10\% corresponds to the most extreme manifestation
of a wider instability (for example, if all stars undergo some
instability in the final burning sequences, but only the most extreme
cases lead to detectably violent mass loss and SNe~IIn) or if SNe IIn
are the outcome of special circumstances in a particular evolution
channel (i.e. interacting binaries within a certain mass and orbital
period range).  There is still very little information available on
any trends with metallicity, although this is always good to
investigate when mass loss plays a critical role.

Recent years have seen something of a paradigm shift in massive star
studies.  Formerly standard or straightforward assumptions about the
simplicity of single-star evolution are giving way to more complicated
scenarios as astronomers grudgingly admit that binary stars not only
exist, but are common and influential (e.g., Sana et al.\ 2012).  This
may be especially true among transient sources and stellar deaths that
seem otherwise peculiar or difficult to understand.  Given the very
high multiplicity fraction among massive stars, binary interaction
should probably not be considered a last resort or the refuge for an
uncreative theorist, but rather, a default assumption.

Whether they are binaries or not, all SNe~IIn require some major shift
in stellar structure and mass loss before the SN.  The synchronization
with core collapse gives a strong implication that something wild is
happening in the latest sequences of nuclear burning.  Those cases
where a star changes its structure as a result of these nuclear
burning instabilities (i.e. an inflated envelope) in a close binary
system seem promising for inducing sudden eruptive behavior for
various reasons (Smith \& Arnett 2014).  Single stars may also be able
to induce their own violent mass loss in the couple years preceding
core collapse due to energy transported to the envelope from Ne and O
burning zones (Quataert \& Shiode 2012).  However, we do not yet have
a good explanation in single-star evolution for the strong mass-loss
that leads to the ``enduring'' class of SNe~IIn or some of the more
extreme cases of ``delayed onset'' of CSM interaction, where the
strong interaction that lasts for years after explosion suggests very
strong mass loss for {\it decades or centuries} before core collapse.
If binary interaction is an important ingredient, then this sort of
interaction before a SN might make asymmetric CSM very common,
suggesting that we should be paying close attention to possible
observed signatures of asymmetry.

\begin{acknowledgement}
  My attempt to understand interacting SNe and their connections to
  massive stars has benefitted greatly from conversations with
  numerous people, but especially Dave Arnett, Matteo Cantiello, Nick
  Chugai, Selma de Mink, Ori Fox, Morgan Fraser, Dan Kasen, Jon
  Mauerhan, Stan Owocki, Jose Prieto, Eliot Quataert, Jorick Vink, and
  Stan Woosley.  While drafting this chapter, I received support from
  NSF grants AST-1312221 and AST-1515559.
\end{acknowledgement}


{\bf Cross-References}

\begin{enumerate}
\item{Supernova of 1054 and its Remnant, the Crab Nebula}
\item{Observational and Physical Classification of Supernovae}
\item{Supernovae from super AGB Stars (8-12 Msun)}
\item{Supernovae from massive Stars (12-100 Msun)}
\item{Very Massive and Supermassive Stars: Evolution and Fate}
\item{Progenitor of SN 1987A}
\item{Supernova Progenitors Observed with HST}
\item{Interacting Supernovae and the Influence on Spectra and Light Curves}
\item{X-ray and radio emission from circumstellar interaction}
\item{Dust and Molecular Formation in Supernovae}
\item{Supernova remnant from SN1987A}
\item{The Supernova - Supernova Remnant Connection}
\end{enumerate}

%
%

%
%

\end{document}